\documentclass[twocolumn,notitlepage,floats,nofootinbib,amsmath,amssymb,aps,prd]{revtex4-1}

\usepackage{amsmath,amssymb,amsfonts,latexsym,cancel,amsthm}

\usepackage{mathtools}
\usepackage{graphicx}
\usepackage{color}
\usepackage{amsbsy}
\usepackage{hyperref}
\usepackage{tikz}

\usepackage{seqsplit}
\usepackage{comment}

\newcommand{\be}{\begin{equation}}
\newcommand{\beq}{\begin{equation}}
\newcommand{\ee}{\end{equation}}

\newcommand{\abs}[1]{\left\lvert#1\right\rvert}

\def\bea {\begin{eqnarray}}
\def\eea {\end{eqnarray}}

\def\dd{{\rm d}}

\definecolor{newgreen}{rgb}{0.0, 0.75, 0.0}
\definecolor{cadmiumgreen}{rgb}{0.0, 0.42, 0.24}

\newtheorem*{theorem}{Theorem}

\begin{document}

\title{Non-uniqueness of the shockwave dynamics in effective loop quantum gravity }

\author{Francesco Fazzini} \email{francesco.fazzini@unb.ca}
\affiliation{Department of Mathematics and Statistics, University of New Brunswick, \\
Fredericton, NB, Canada E3B 5A3}

\begin{abstract}

Spherically symmetric effective dust collapse inspired by effective loop quantum cosmology predicts a bounce when the stellar energy density becomes planckian, which in turn inevitably leads to shell-crossing singularity formation. An extension of the spacetime beyond such singularities is possible through weak solutions of the equations of motion in integral form, leading to the shockwave model. In this work, we show explicitly that such an extension is not unique, and that relevant features like the black hole life-time strongly depend on the choice of the integral form of the equation of motion.

\end{abstract}

\maketitle

\section{Introduction}

The final fate of black holes is a central problem in any approach of quantum gravity. In the classical framework, once a star collapses to form a black hole, the collapse continues until it  reaches the so called shell-focusing singularity (SFS), according to Penrose singularity theorem \cite{Penrose:1964wq}. However, once loop quantum gravity (LQG) effects are kept into account, most models constructed so far lead to the resolution of the singularity and its replacement by a quantum gravitational bounce \cite{Rovelli:2014cta,Husain:2021ojz,Han:2023wxg,Giesel:2023hys,Lewandowski:2022zce,Fazzini:2023scu,Husain:2022gwp,Kelly:2020lec,BenAchour:2020gon,Giesel:2021dug,Olmedo:2017lvt,Bianchi:2018mml,Christodoulou:2016vny,Munch:2020czs,BenAchour:2020mgu,Bobula:2023kbo, Cafaro:2024vrw, Alonso-Bardaji:2023qgu,Bonanno:2023rzk,BenAchour:2020bdt,Liu:2014kra,Bambi:2013caa}. The bounce arises when the energy density reaches the Planck scale for any initial energy density profile, both in the case of dust and stars with pressure \cite{Cafaro:2024lre}.\par
In order to modify Einstein's field equations to include loop quantum gravity corrections at the effective level, one needs to make assumptions. These choices primarily concern variables to be polymerized, the choice of the holonomy length, the requirement of classical or deformed covariance, the inclusion or exclusion of inverse triad corrections, and which constraint needs to be polymerized. A common feature of effective models in loop quantum gravity is instead the reduction to spherical symmetry before polymerization.

The results presented are based on the so called 'K-loop quantization' as polymerization scheme \cite{Vandersloot:2006ws,Singh:2013ava}, the $\Bar{\mu}$ scheme for the holonomy length \cite{Ashtekar:2006wn,Boehmer:2007ket,Gambini:2020qhx,Kelly:2020uwj} and the neglect of inverse triad corrections. The diffeomorphism constraint is not polymerized \cite{Giesel:2023hys,Giesel:2023tsj}, while the dust-time gauge is fixed at the classical level  \cite{Husain:2021ojz,Husain:2022gwp,Giesel:2023hys,Giesel:2023tsj}. The resulting effective model reproduces standard effective loop quantum cosmology (LQC) in its cosmological sector \cite{Ashtekar:2006wn}, and has been extensively studied in the dust case \cite{Husain:2021ojz,Husain:2022gwp,Fazzini:2023ova,Fazzini:2023scu,Cafaro:2024vrw,Cipriani:2024nhx,Munch:2020czs,Munch:2021oqn,Giesel:2023tsj,Lewandowski:2022zce}, and recently generalized to fluids with pressure \cite{Wilson-Ewing:2024uad,Cafaro:2024lre}. 

A key feature of inhomogeneous collapse within this model is the development of shell-crossing singularities (SCS) within a time $\delta t =(2/3)\sqrt{\Delta}$ after the quantum gravitational bounce \cite{Fazzini:2023ova}. Specifically, SCS arise for any initial continuous profile of compact support, and for non-compact inhomogeneous continuous profiles with sufficiently large inhomogeneities \cite{Fazzini:2023ova}. These results were derived using the Lemaître-Tolman-Bondi (LTB) gauge, which is well-suited for studying SCS formation. 

Once SCS form, the energy density develops a divergence which, in turn, produces a divergence in curvature scalars (see \cite{Hellaby:1985zz,Szekeres:1995gy} for an analysis of SCS in the classical context). These singularities, despite being physical, are called \emph{weak}, as opposed to the classical SFS. In contrast to SFS, tidal forces do not diverge close to the singular spherical surface \cite{Szekeres:1995gy}.

As observed in the classical case, even if the equations of motion cease to hold when SCS form in any gauge (due to the physical character of the singularity), it is possible, in principle, to extend the spacetime dynamics beyond SCS \cite{Nolan:2003wp}. This is explicitly realized by looking at the integral form of the equations of motion (EoMs). The mathematical motivation behind this approach stems from the fact that the classical equation of motion for dust collapse in the Painlevé-Gullstrand (PG) gauge is a non-linear hyperbolic conservation law, which commonly develops discontinuities in the solution  \cite{Leveque:2002}. This kind of differential equation arises in many areas of physics, including classical fluid dynamics \cite{article,Batchelor_2000}, quantum fluid dynamics \cite{Hoefer_2006}, and supernovae physics \cite{Chavanis:2019auu}. The integral approach turned out to be successful in capturing the relevant physics of these problems.

The integral approach has been applied also to effective dust collapse within the model considered in this paper (\cite{Husain:2022gwp} for the marginally bound case, \cite{Cipriani:2024nhx} for the non-marginally bound case). The resulting dynamics consists in an outgoing propagating thin shell (a dynamical shell-crossing singularity) located at any time on a discontinuity of the extrinsic curvature. This gravitational shockwave, emerging from the trapped region of the black hole, leads to a black hole "explosion" when its location reaches the outer black hole horizon. The predicted black hole lifetime is proportional to $M^2$, where $M$ is the gravitational mass of the collapsed star. 

While the integral form of the equations of motion allows extending the dynamics beyond SCS (or, in PDE language, physical characteristic crossings), it introduces significant drawbacks. Unlike the original PDE, the dynamics derived from this approach is not unique, and such non-uniqueness is two-fold \cite{Leveque:2002}: firstly, the integral form of the equation is non-unique, in the sense that different integral forms of equivalent (but distinct) PDEs yield different weak solutions after characteristic crossing. Secondly, even from the same integral equation, different weak solutions can arise from a single initial condition. We emphasize here that such non-uniqueness is a standard result in fluid dynamics (and other fields that study weak solutions).

The second issue is the less critical, as \emph{entropy conditions} exist to select the physically meaningful weak solution and discard others \cite{Leveque:2002}. As a side note, it can be easily verified that the effective shockwave dynamics described before satisfies such conditions. However, no mathematical method addresses the first issue (non-uniqueness of integral forms) which is the focus of this work.

The present study will explicitly demonstrate how different equations in integral form, derived from the same PDE, lead to different shockwave dynamics. In particular, an infinite class of such equations will be derived, each connected to a corresponding black hole lifetime. This result, foreshadowed conceptually in \cite{Fazzini:2023ova}, will be explicitly derived here, and it will allow interpreting the shock dynamics presented in \cite{Husain:2022gwp} as an element of an infinite class of gravitational shockwave dynamics that are compatible with the same underlying effective PDE.

\section{The class of equations of motion\label{section2}}

The starting point of our analysis is the PDE governing effective dust collapse in the $\Bar{\mu}$ + $K$-loop quantization scheme. In the marginally bound case, it reads \cite{Husain:2022gwp}
\begin{equation}
 \partial_t b(r,t)+ \frac{1}{2\gamma \Delta r}\partial_{r}\left[{r^3}\sin^2\left(\frac{ \sqrt{\Delta} b}{r} \right) \right]=0 ~,  \label{pde}
\end{equation}

where $\Delta \sim l_{P}^2$ is the minimum area gap in LQG, $\gamma$ is the Babero-Immirzi parameter, and $b$ in the classical limit is the angular component of the extrinsic curvature. The equation is written in Painlevé-Gullstrand (PG) coordinates, therefore $r$ is the areal radius and $t$ is the advanced PG time, i.e. the proper time of radial free-falling time-like observers. The line element in these coordinates takes the form
\begin{equation}
\dd s^2=-\dd t^2  + (\dd r + N^r \dd t)^2 +r^2 \dd \Omega^2 ~,
\end{equation}
where $N^r$, the radial component of the shift vector, is given by $N^{r}=-\frac{r}{\gamma \sqrt{\Delta}}\sin\left(\frac{\sqrt{\Delta}b}{r} \right) \cos\left(\frac{\sqrt{\Delta}b}{r} \right)$ \cite{Husain:2022gwp,Giesel:2023hys}.

In order to simplify \eqref{pde}, one can make the change of variables $B=r\sqrt{\Delta}b $, yielding
\begin{equation}
\partial_t B(r,t)+\partial_r \left[\frac{r^3}{2 \gamma \sqrt{\Delta}}\sin^2\left(\frac{ \sqrt{B}}{r^2} \right) \right]=0~.
\label{pde1}
\end{equation}
The basic idea for constructing  integral equations different from the trivial one (obtained by simply integrating \eqref{pde1} in $r$ \cite{Husain:2022gwp}), is to multiply \eqref{pde1} by functions $f(B(r,t),r)$, and then rearrange the result to regain a conservation law of kind 
\begin{equation}
  \partial_t v(r,t) +\partial_r f(v(r,t),r) =0  ~.
  \label{proto}
\end{equation}
To achieve this goal, we multiply \eqref{pde1} by $r^{\frac{3n}{2}}\sin^{n}\left(\frac{B}{r^2} \right)$ (with $n \in \mathbb{N}$),
and rearrange the last term, yielding
\begin{equation}
\sin^n \left( \frac{B}{r^2}\right)\partial_t ( r^{\frac{3n}{2}}B) +\frac{1}{(n+2)\gamma \sqrt{\Delta}}\partial_r\left[r^{3}\sin^2 \left(\frac{B}{r^2} \right)\right]^{\frac{n}{2}+1}  =0 ~. 
\label{eqn}
\end{equation}

In order to write the first term in the required form \eqref{proto}, we use the following identity:
\begin{equation}
\sin^n\left( \frac{B}{r^{2}}\right)=\sum_{k=0}^{n}\frac{(-1)^k}{2^n}\binom{n}{k}\cos\left[\frac{(n-2k)B}{r^{2}}-\frac{\pi}{2}n\right]  ~.
\label{class0}
\end{equation}
It is then straightforward to check that
\begin{align}
&\sin^{n}\left( \frac{B}{r^{2}}\right)\partial_t ( r^{\frac{3n}{2}}B)= \notag \\
&=\partial_t\left\{r^{2+\frac{3}{2}n}\sum_{k=0}^{n}\frac{(-1)^k}{2^n(n-2k)}\binom{n}{k}\sin\left[\frac{(n-2k)B}{r^{2}}-\frac{\pi}{2}n\right]\right\} ~.  
\label{class}
\end{align}
This, combined with \eqref{eqn} and \eqref{class0} gives an infinite class equations of kind
\begin{equation}
 \partial_t v(B(r,t),r)+ \partial_r f(B(r,t),r)=0 ~,   \label{general}
\end{equation}
parametrized by $n$, where:
\begin{align}
        &  v=r^{2+\frac{3}{2}n}\sum_{k=0}^{n}\frac{(-1)^k}{2^n(n-2k)}\binom{n}{k}\sin\left[\frac{(n-2k)B}{r^{2}}-\frac{\pi}{2}n\right] \label{v}\\
        &f=\frac{r^{3+\frac{3n}{2}}}{(n+2)\gamma \sqrt{\Delta}}\sum_{k=0}^{n+2}\frac{(-1)^{k+1}}{2^{n+2}}\binom{n+2}{k}\cos\left[\frac{(n+2-2k)B}{r^{2}}-\frac{\pi}{2}n\right]  \label{g}
\end{align}
To obtain \eqref{proto}, one must invert \eqref{v} to get a function $B=B(v,r)$ and substitute this result into \eqref{g}. It can be easily shown that \eqref{v}, for any $n\geq 0$, is monotonic (and thus invertible) within the range ${B}/{r^2}\in [-\pi, 0]$. As we will demonstrate later, this range is consistent with the dynamics, allowing to cast the class \eqref{eqn} in the form of non-linear conservation laws. 

It is important to highlight that these equations produce precisely the same dynamics as \eqref{pde1} until characteristics cross (or in general relativistic language SCS arise) and the fields become multi-valued. However, as explicitly shown in the subsequent section, once characteristics cross, the weak solutions of the integral form of these equations differ. This, as previously noted, is a well-known property of general non-linear conservation laws and has been explicitly shown in the classical case \cite{Nolan:2003wp}, where two different forms of the classical limit of \eqref{pde} are considered for the weak dynamics. Furthermore, eq. \eqref{general} generally does not encompass the entire class of equivalent PDEs, although exploring other possible forms is beyond the scope of this work.

\section{Shell-crossing singularities after the bounce}

To analytically investigate the weak solutions generated by our class of PDEs, we limit our analysis to initial data that produce SCS only after the star core's bounce. We therefore construct a theorem that identifies the class of initial profiles fulfilling this requirement. \par
The PDE \eqref{pde1} takes in LTB coordinates the following form \cite{Giesel:2023hys}:
\begin{equation}
    \left( \frac{\Dot{r}}{r}\right)^2=\frac{2G m(R)}{r} \left(1-\frac{2G m(R)\gamma^2 \Delta }{r^3} \right) ~,
    \label{beautiful}
\end{equation}
where $r(R,t)$ is the areal radius of the shell $R$ at time $t$, and $m(R)$ is the (conserved) gravitational mass within $[0,R $]. The analytic solution of the previous equation reads
\begin{equation}
r(R,t)=\left[2Gm(R)\right]^{\frac{1}{3}}\left\{\left[t-\alpha(R)\right]^{2} +\Delta\right\}^{\frac{1}{3}}  ~,
\label{beauty}
\end{equation}
where $\alpha(R)$ is fixed by the initial condition. A fundamental feature of this generic solution, valid for any initial energy density profile, is that each shell $R$ will bounce at $t_{B}(R)=\alpha(R)$. In the pre-bounce (post-bounce) phase we have $t<\alpha(R)$ ($t>\alpha(R)$).

The dust energy density in LTB coordinates reads \cite{Giesel:2023hys,Fazzini:2023ova}
\begin{equation}
 \rho(R,t)=\frac{\partial_R m }{4 \pi r^2 \partial_R r}   
\end{equation}
and shell-crossing singularities arise when the following conditions are both satisfied: $\partial_R r=0$ and $\partial_R m(R) \neq 0$. Since for regular matter ($\rho>0$) inside the star we have $\partial_R m(R) >0$, the only condition needed for SCS formation is $\partial_R r(R,t)=0$. By applying the spatial derivative on the solution \eqref{beauty}, we get
\begin{equation}
 \frac{\partial_{R}r}{r}=\frac{\partial_{R}m}{3 m}+\frac{3}{2}\frac{(\alpha-t)\partial_{R}\alpha}{[\frac{9}{4}(\alpha-t)^{2}+\Delta]} ~.
 \label{crucial}
 \end{equation}
We have all the ingredients at this point to prove the following
\begin{theorem}
For the case $\mathcal{E} = 0$, a shell-crossing singularity forms inside the star within a time $t_{B}(R)<t<t_{B}(R)+(2/3)\sqrt{\Delta}~,$ if $\partial_R \rho(t_0,R)<0$ and the initial condition satisfies $\frac{m\partial_R \alpha}{\partial_R m}> (2/3)\sqrt{\Delta}$.
\end{theorem}
\noindent \emph{Proof:} The first part of the proof aims to prove that  $\partial_R \rho(R,t_0)<0$ implies $\partial_R \alpha>0$. We show this by considering as initial condition for $r$: $r(t_{0},R)=R$.  
 By inverting \eqref{beauty}, and evaluating it at initial time it is straightforward to check that for an initial collapsing profile \cite{Fazzini:2023ova}
\begin{equation}
 \alpha(R)=t_0+\sqrt{\frac{4}{9}\left(\frac{R^3}{2G m(R)}-\Delta \right)   } ~.
\end{equation}
By differentiating the previous expression with respect to $R$, we obtain
\begin{equation}
 \partial_{R}\alpha=\frac{R^2}{2G m(R) \sqrt{\frac{R^3}{2Gm }-\Delta}}\left(1-\frac{R \partial_{R}m}{3m} \right) ~.  
 \label{derivativea}
\end{equation}
Now, let us consider an initial decreasing profile $\partial_R \rho(t_0,R)<0$, for any $R$. We can write
\begin{equation}
R^2\rho(t_0,R)> \frac{\partial_{R}\rho(t_0,R)R^3}{3}+R^2 \rho(t_{0},R)    
\end{equation}
for any $R>0$. This means
\begin{equation}
 R^2 \rho(t_0,R)> \partial_R \left[\frac{\rho(t_0,R)R^3}{3}  \right] ~.
 \label{derivative}
\end{equation}
Now, let us integrate both the members

\begin{equation}
\int_{0}^{R} \rho(t_{0},R')R'^2\dd R'> \rho(t_{0},R) \frac{R^3}{3}.    
\label{inequality}
\end{equation}
This last inequality holds for any $R>0$, since \eqref{derivative} and the fact that the left and right of \eqref{inequality} are zero at $R=0$. The previous relation can be written as
\begin{equation}
 m(R)>\frac{1}{3} R \partial_R m(R) ~.  
 \label{impo}
\end{equation}
By plugging this inequality in \eqref{derivativea} we obtain $\partial_R \alpha>0$. This implies, looking at \eqref{crucial}, that during the pre-bounce phase $r'>0$ for any $R$, and SCS do not arise. By the other side, at time $t=(2/3)\sqrt{\Delta} +t_{B}$, we get
\begin{equation}
 \frac{\partial_{R}r}{r}=\frac{\partial_{R}m}{3 m}-\frac{\partial_{R}\alpha}{2\sqrt{\Delta}}~.
 \end{equation}
Therefore, if $\frac{m\partial_R \alpha}{\partial_R m}> (2/3)\sqrt{\Delta}$ we get $r'(R, t_{B}+\frac{2}{3}\sqrt{\Delta})<0$, and since $r'(R,t)$ is a continuous function, there will be some time $t_{B}(R)<t<t_{B}(R)+(2/3)\sqrt{\Delta}$ for which $r'=0$, that implies SCS formation for a shell $R$ inside the star ($\partial_R m(R)>0$). $\Box$ \par
The condition $\frac{m\partial_R \alpha}{\partial_R m}> (2/3)\sqrt{\Delta}$, as proved in \cite{Fazzini:2023ova}, can be translated in a condition on the degree of inhomogeneity of the initial profile: for initial decreasing continuous profiles (not necessarily of compact support) for which $\partial_R \rho(R,t_0)$ is large enough, this condition is satisfied. Moreover, as emphasized in \cite{Fazzini:2023ova}, it is extremely easy to find initial profiles that fulfill such condition. \par
This result, valid for both \eqref{pde1} and \eqref{general}, allows to safely assume that characteristic crossings arise only in the post-bounce dynamics for decreasing initial density profiles inhomogeneous enough.

\section{General features of the weak solutions}

As previously stated, the solutions to equations \eqref{pde1} and \eqref{general} are identical until characteristics cross (i.e., a shell-crossing singularity arises). When characteristics cross, one must consider the integral form of the equations and analyze the dynamics of the resulting weak solution. Note that characteristic crossings can occur in vacuum or between a physical shell and a vacuum shell (as in Oppenheimer-Snyder collapse). In these cases, since no physical singularity develops, the PDE breakdown is a coordinate artifact and can be eliminated through a coordinate transformation. Consequently, introducing weak solutions is unjustified \cite{Fazzini:2023scu,Giesel:2023hys}.

Regardless of the particular form of the integral equation, weak solutions remain equivalent to the solution of \eqref{pde1} (or \eqref{beautiful} in LTB coordinates), until shell-crossing singularities arise. At this point, while the differential form of the equations is no longer due to multi-valued solutions, the weak solution continues its evolution, treating the characteristic crossing as a discontinuity in the field variables. Notably, as shown in \cite{Husain:2022gwp} and validated numerically, the weak solution keeps agreeing with the differential form solution outside the discontinuity region. Therefore, given that the PDE in the differential form holds before the bounce and far from SCS, this solution can be used to describe spacetime evolution inside and outside the shockwave also for weak solutions derived from the integral form of \eqref{general}, as done in \cite{Husain:2022gwp} for the particular $n=0$ case.

If we further consider a stellar energy density profile that at initial time is almost flat in the interior, with a sharp tail and of compact support, we can approximate the post-SCS dynamics for the interior as an effective Friedmann dynamics, and outside as the effective vacuum  \cite{Kelly:2020uwj}. This approximation turned out to be highly accurate for profiles with sharp tails (\cite{Husain:2022gwp} for the marginally bound case, \cite{Cipriani:2024nhx} for the non-marginally bound case). Since weak solutions are equivalent in regions where the solution of the PDE is valid, this equivalence should hold also for weak solutions derived from the integral form of \eqref{general}.
 
This implies that the field $B(t,r)/r^{2}$ after both the bounce and SCS formation can be approximated with \cite{Husain:2022gwp} 
\begin{align}
\frac{B(t,r)}{r^2} \sim
\begin{cases}
 &  -\pi+\arcsin\frac{1}{\sqrt{1+\frac{9t^2}{4\gamma^2 \Delta}}}   ~, ~ r\leq L(t) \\
 &-\arcsin \sqrt{\frac{\gamma^2 \Delta R_S}{r^3}}   ~, ~ r>L(t)
 \end{cases}
 \label{system}
\end{align}
where $R_S=2GM$ is the classical Schwarzschild radius of the star with total gravitational mass $M$, and $L(t)$ is the location of the shock at time $t$. Moreover, for $t\gg \frac{2\gamma\sqrt{\Delta}}{3}$ (non-planckian time after the bounce), $B(t,r)\sim -\pi r^{2}$ for $r\leq L(t)$. Notice that for this solution the bounce happens at $t=0$ \cite{Husain:2022gwp}, according with \eqref{beauty} in the homogeneous cosmological case, where $\alpha(R)=0$ (see \cite{Giesel:2023hys} for an analysis of the cosmological solution). This comes from having required that the interior is almost homogeneous. 

 Since outside the shock the weak solutions of the integral form of any equation of the class \eqref{general} must agree with \eqref{system}, we can safely assume that the $B/r^{2}$ field during the shock dynamics takes values in $[-\pi, 0]$ for such weak solutions. Since the same holds also for the pre-bounce dynamics \cite{Husain:2022gwp}, and this range is the one in which the function $v(B,r)$ is invertible, the equations \eqref{general} can be cast in the form \eqref{proto}, and there is no invertibility issue for any $n \in \mathbb{N}$. Even if finding the inverse of \eqref{v} for generic $n$ is a non-trivial task, it is outside the aim of this work since not necessary to derive the analytic shockwave dynamics. Ti is however a crucial step for the construction of the numerical scheme, and will be a subject of future work.

\section{Shockwave velocities and black hole lifetimes}
 
What is left is evaluating the shock velocity $\dd L (t) / \dd t$. 
This can be done analytically by looking at the Rankine-Hugoniot condition (see \cite{Leveque:2002} for a general treatment). Such a condition comes from integrating the PDE over $r$ within an interval $r\in [r_{1},r_{2}]$ that is assumed to contain the shock at time $t$, and then making the limit $r_{1,2}\rightarrow L_{\mp}$  \cite{Husain:2022gwp}. The result, for a generic conservation law of kind \eqref{proto}, is
\begin{equation}
 \frac{\dd L(t)}{\dd t}=\frac{\left[ f \right]_{-}^{+}}{\left[v\right]_{-}^{+}}  ~~,
 \label{shockvel}
\end{equation}
where $[A]^{+}_{-}\equiv lim_{r\rightarrow L^{+}}A(r)\equiv lim_{r\rightarrow L^{-}}A(r)$. We are now in the position of computing the shock velocity for the integral form of the class of PDEs \eqref{general}. By using \eqref{shockvel}, we get 
\begin{equation}
 \Dot{L}=\frac{ 2^n \left[\abs{\sin^2\left(\frac{B}{L^2} \right)}^{\frac{n}{2}+1}\right]^{+}_{-} L }{ \left[\sum_{k=0}^{n}{(-1)^k}\binom{n}{k}\sin\left[(n-2k)\frac{B}{L^2}-\frac{\pi n}{2} \right]\frac{\gamma\sqrt{\Delta}(n+2)}{(n-2k)}    \right]^+_-}   ~.
\end{equation}
Notice that this expression is written in terms of the variable $B$. By substituting \eqref{system} in the numerator, we get
\begin{equation}
\Dot{L}=\frac{(-1)^n\gamma^{n+1} 2^n \Delta^{\frac{1}{2}(n+1)}R_S^{\frac{n}{2}+1}L^{-\frac{3}{2}n-2}}{(n+2)\left[\sum_{k=0}^{n}{(-1)^k}\binom{n}{k}\sin\left[(n-2k)\frac{B}{L^2}-\frac{\pi n}{2} \right]\frac{1}{(n-2k)}    \right]^+_-} ~.
\end{equation}
In order to write the denominator explicitly, we need to distinguish between odd and even $n$.
\subsubsection{Shock velocity and black hole lifetime for n even}
In the case of $n$ even, it is important noticing that the $k=n/2$ term in the series is actually given by:
\begin{align}
a_{k=n/2}=&(-1)^{\frac{n}{2}}\binom{n}{n/2}\lim_{k \rightarrow n/2} \frac{\sin\left[(n-2k)\frac{B}{L^2}-\frac{\pi n}{2} \right]}{n-2k}=\notag \\
=& \binom{n}{n/2}\frac{B}{L^2}.
\end{align}
Let's look first at the denominator computed in the interior ($-$ term). Since $B_{int}(r,t)=-\pi L^2$, it is easily to see that all the terms in the series are zero, except for the $k=\frac{n}{2}$ term, that we call $a^{-}_{k=\frac{n}{2}}$. It can also be easily verified that for $L\gg(\gamma^{2} \Delta R_S )^{\frac{1}{3}}$ (that holds suddenly after the bounce), the series in the $+$ term gives contributions much smaller than $a^{-}_{k=n/2}$. Therefore, suddenly after the shockwave formation, for $n$ even we get
\begin{equation}
\Dot{L}\sim \frac{\gamma^{n+1}2^n \Delta^{\frac{n+1}{2}}R_{S}^{\frac{n}{2}+1}}{\pi (n+2)L^{\frac{3}{2}n+2} \binom{n}{n/2}}  ~.
\end{equation}
The black hole lifetime can be easily derived solving the previous ODE, getting
\begin{equation}
T_{BH}(R_s |n)\sim \frac{2\pi R_S^{n+2}}{3\gamma^{n+1}2^n \Delta^{\frac{n+1}{2}} }  \binom{n}{n/2}
\end{equation}
which for the particular case $n=0$ reduces to the result of \cite{Husain:2022gwp}: $T_{BH} \sim \frac{2 \pi R_S^2}{\gamma \sqrt{\Delta}}$. Therefore, the black hole lifetime for even $n$ is proportional to even powers of $R_S$.
In the ODE solution we assumed: $t_0=0$, $L(t_0)=0$, $t_f \equiv T_{BH}$, $L(t_f)=R_S$. Indeed, the black hole life-time for the $n=0$ case can be safely approximated with the time required to the shock to exit the black hole outer horizon \cite{Husain:2022gwp}, and $T_{BH}(R_S|n)>T_{BH}(R_S|0)$ for macroscopic black holes.

\subsubsection{Shock velocity for $n$ odd}
In the case of $n$ odd the numerator for the shock velocity is the same as before, with the exception of a $-1$ pre-factor. Let's look therefore at the denominator, in particular at the $-$ term and call it $D^-$. Keeping in mind that $B_{-}\sim -\pi L^2$, one can use the trigonometric duplication formula for the $\sin$ function getting:
\begin{equation}
D^{-}=(n+2)L^{\frac{3}{2}n+2}\sum_{k=0}^{n}(-1)^{k}\binom{n}{k}\frac{1}{n-k}~. 
\end{equation}
Let's finally look at $D^+$. Firstly, we notice from the second of \eqref{system}, that the "angle" $B/r^2$ is in the fourth quadrant, that means $\cos\left(\frac{B^+}{L^2} \right)=\sqrt{1-\frac{\gamma^2 \Delta R_S}{L^3}}$, taken with the plus sign. Then, for $L^3>> \gamma^2 \Delta R_S$ (soon after the bounce, we can approximate the previous expression to 
\begin{equation}
\cos\left(\frac{B^+}{L^2} \right)\sim 1-\frac{\gamma^2 \Delta R_S}{2 L^3} ~.
\end{equation}
By using the duplication formula for $\sin$, combined with the previous result, we find:
\begin{equation}
D^+=L^{\frac{3}{2}n+2} \sum_{k=0}^{n} (-1)^{k+n} \binom{n}{k} \frac{(n+2)}{n-2k} \left[1-(n-2k) \frac{B^+}{L^2} \right]  ~, 
\end{equation}
and, collecting the terms, we get
\begin{equation}
\Dot{L}=\frac{\gamma^{n+1} 2^n \Delta^{\frac{1}{2}(n+1)}R_S^{\frac{n}{2}+1}L^{-\frac{3}{2}n-2}}{(n+2)\sum_{k=0}^{n}{(-1)^{k}}\binom{n}{k}\left[ 2 -(n-2k)\frac{\gamma^2 \Delta R_S}{L^3} \right]\frac{1}{(n-2k)}} ~,
\end{equation}
which, soon after the bounce, can be safely approximated to
\begin{equation}
\Dot{L}\sim \frac{\gamma^{n+1} 2^{n-1} \Delta^{\frac{1}{2}(n+1)}R_S^{\frac{n}{2}+1}L^{-\frac{3}{2}n-2}}{(n+2)\sum_{k=0}^{n}{(-1)^k}\binom{n}{k}\frac{1}{(n-2k)}}   ~. 
\end{equation}
With this in hand, one can compute the black hole lifetime for odd $n$, by solving the previous ODE. The result
\begin{equation}
T_{BH}(R_S|n)\sim \frac{2 R_S^{n+2} \sum_{k=0}^{n}(-1)^k \binom{n}{k} \frac{1}{n-2k}}{3\gamma^{n+1} 2^{n-1} \Delta^{\frac{n+1}{2}}}    
\end{equation}

is a lifetime with odd powers of the black hole mass. Notice that for $n=1$ it reduces to $T_{BH}=\frac{4}{3}\frac{R_S^3}{\gamma^2\Delta}$, which is proportional to the black hole evaporation time. 

To conclude, it is possible in principle to find weak solutions related to \eqref{eqn} for $n<0$. This would imply presumably a life-time shorter than $M^2$. We do not investigate this possibility here since, except for \eqref{system}, the other approximations carried for the $n \geq 0$ case cannot be applied, and numerical investigation is necessary.

\section{Astrophysical implications and relation with other models}

An appealing aspect of the shockwave model presented in \cite{Husain:2022gwp} is its potential avoidance of the information paradox. The information carried by the collapsed star would emerge from the black hole's outer horizon within a time $T\propto M^2$, which is shorter than the Page time. Furthermore, if the model is at least qualitatively correct, such black hole "explosions" could leave observable astrophysical signatures.

However, the non-uniqueness of the shockwave dynamics presents a significant mathematical challenge, anticipated at the conceptual level in \cite{Fazzini:2023ova}, and typical of weak solutions of non-linear conservation laws.
Importantly, even though the transformation \eqref{eqn} does not apply to discontinuous solutions, there is no mathematical prescription to start with \eqref{pde1} instead of \eqref{general}. Despite this, from a physical perspective, this ambiguity could enlarge the window of quantum gravitational signatures from black hole physics. Just as we have found dynamical extensions that predict a lifetime greater than $M^2$, it is in principle possible to find extensions that predict a shorter black hole lifetime. This suggests a possible avenue for parameter-dependent lifetimes, potentially constrained by astrophysical observations. Models with $T\ll M^2$ could predict a much higher number of exploded black holes throughout the universe's history than the $n=0$ case explored in \cite{Husain:2022gwp, Cipriani:2024nhx}, increasing the chances of astrophysical detection.

Another possible scenario needs consideration. Numerical simulations of the $n=0$ case show that during the post-bounce dynamics, the outward motion of the shockwave does not allow sufficient time for the external vacuum to bounce and form an anti-trapped region, unlike what occurs in the Oppenheimer-Snyder case (see e.g. \cite{Lewandowski:2022zce,Giesel:2023hys, Fazzini:2023scu, Munch:2021oqn, Bobula:2023kbo}), where the vacuum trapped region has time to bounce and evolve in an non-trapped one. This leads to a shockwave moving toward the trapped region that is super-luminal until the shock reaches the outer horizon, if computed with the external vacuum metric. Within the class of equations \eqref{eqn}, those that predict a longer $T_{BH}$ might permit the vacuum to bounce, forming a white hole anti-trapped region and causing the shockwave to emerge in a second asymptotic region. This could lead to dynamics compatible with \cite{Lewandowski:2022zce,Giesel:2023hys, Fazzini:2023scu, Munch:2021oqn, Bobula:2023kbo}, or align with
the black-to-white hole transition picture \cite{Han:2023wxg}. If this is the case, the astrophysical implications could dramatically differ from the shockwave model proposed in \cite{Husain:2022gwp, Cipriani:2024nhx}. In the first scenario, the shockwave might emerge from another universe, leaving our universe with an almost classical evaporating black hole, posing significant challenges to the avoidance of the information paradox. In the second scenario, a shockwave emerging from a white hole remnant with planckian mass could represent the final stage of black hole evaporation. Further investigation into the concrete connection between our results and these scenarios is left for future research.

Constraints on the black hole lifetime can arise from theoretical or experimental grounds. For example, it cannot be too small, say $\leq M$, since, as pointed out in \cite{Ashtekar:2023cod}, these possibilities are ruled out by experimental data. From the theoretical side, one could require sub-luminal motion of the shockwave dynamics, as measured by \emph{any} metric, which would rule out at least the $\leq M^2$ models. Oppositely, if one expects that a quantum gravitational star collapse model should solve issues arising from less fundamental approaches, like the information paradox, without ad-hoc assumptions already at the effective level, the life-time allowed should lie in the range between $M$ and $M^2$. 
The ambiguity could be fixed if one is able to find variables (if any) that do not develop discontinuities during the evolution, avoiding to seek for weak solutions.
Also, it could be overcome looking at a more fundamental approach, like group field theory \cite{Oriti:2014uga}, where one can control both the effective dynamics and microscopic features like the types of interactions between quanta of space. Different effective outcomes could then be constrained (or characterized) by different kind of interactions, or the ambiguity completely avoided if the effective equations turn out to not develop shell-crossing singularities, through the inclusion of other quantum gravitational effects. Further work is needed to investigate these scenarios.

\acknowledgments

I would like to thank Edward Wilson-Ewing, Viqar Husain and Luca Cafaro for helpful comments. This work is supported in part by the Natural Sciences and Engineering Research Council of Canada.

\end{document}